\documentclass[10pt,twocolumn]{article}

\usepackage{cite} 
\usepackage{url}  
\usepackage{multicol}   

\usepackage{graphicx}
\usepackage{tabularx}

\begin{document}
\onecolumn
\title{Biological Profiling of Gene Groups utilizing Gene
Ontology}

\author{
        Nils Bl\"uthgen$^{1}$%
      \and
        Karsten Brand$^2$%
      \and
        Branka \v{C}ajavec$^{1}$%
      \and
        Maciej Swat$^{1}$%
      \and
        Hanspeter Herzel$^{1}$%
      \and 
        Dieter Beule$^3$%
      }

\maketitle
      
\thanks{
    $^1$ Humboldt University, Institute for Theoretical Biology, Invalidenstra\ss e 43, 10115 Berlin, Germany \\
    $^2$Humboldt University, Department of Biology,  Max-Delbr\"uck-Centrum, Robert-R\"ossle-Stra\ss e 10, 13125 Berlin, Germany\\
    $^3$ MicroDiscovery GmbH, Marienburger Stra\ss e 1, 10405 Berlin, Germany
}%

\begin{abstract}
Increasingly used high throughput experimental techniques, like DNA or
protein microarrays give as a result groups of interesting, e.g.\
differentially regulated genes which require further biological
interpretation. With the systematic functional annotation provided by
the Gene Ontology the information required to automate the 
interpretation task is now accessible. However, the determination of
statistical significant e.g.\ molecular functions within 
these groups is still an open question. In answering this question,
multiple testing issues must be taken into account to avoid misleading
results. Here we present a statistical framework that tests whether
functions, processes or locations described in the Gene Ontology are
significantly enriched within a group of interesting genes when
compared to a reference group. First we define an exact analytical
expression for the expected number of false positives that allows us
to calculate adjusted p-values to control the false discovery
rate. Next, we demonstrate and discuss the capabilities of our
approach using publicly available microarray data on cell-cycle
regulated genes. Further, we analyze the robustness of our framework
with respect to the exact gene group composition and compare the
performance with earlier approaches. The software package GOSSIP
implements our method and is made freely available at \url{http://gossip.gene-groups.net/}.
\end{abstract}
\twocolumn
\section{Introduction} 
With the advent of genome-wide screening experiments, like microarray
studies \cite{Duggan99}, and 2D protein gel analyses \cite{Klose02},
researchers frequently face the task of interpreting the biological
function and relevance of gene groups, e.g.\ groups of differentially
expressed genes. Even if the individual genes are annotated this
interpretation task remains laborious and complex. Analysis of gene
groups using an ontology is a promising starting point for an
automated biological profiling beyond the single-gene level. An
ontology specifies a controlled vocabulary and the relations between
the terms within the vocabulary. This concept is widely used to
systematically represent knowledge for further analysis. For molecular
biology the Gene Ontology Consortium provides the Gene Ontology (GO)
as an international standard to annotate functions, affiliation with
processes and locations of genes and gene products \cite{Ashburner00}.

We utilize GO to test whether a molecular function, biological
process, or cellular location (which we call a {\em term} in the
following) is significantly associated with a group of interesting
genes. The definition of statistical significance is a major challenge
due to the large number of terms which need to be tested. The use of
single test p-values is only justified if we test whether a single
term is associated with a specific gene group. However, in genome-wide
screening experiments the situation is fundamentally different: the
current GO includes over 17700 terms, out of which typically several
thousand terms appear in the annotation of an investigated gene group
and which have to be tested. If one performs that many tests, problems
arising from multiple testing cannot be left aside. Namely, even when
we apply a very conservative threshold like $p<0.001$ a few terms will
be reported to be associated with the test group by sheer chance. This
phenomenon is known as false positives or type-I-error. The standard
solution for this problem is to calculate adjusted p-values. These
adjusted p-values control the number of false discoveries in the
entire list and can be used similar to normal p-values for single
tests \cite{Dudoit03}. There are several standard methods to calculate
adjusted p-values, like resampling and multi-step estimations. We find
that adjusted p-values obtained with standard multiple testing
correction methods are unsatisfactory because they are either not
precise enough, or they require major computational efforts.

In this article, we present a statistical framework and the software
GOSSIP to cope with this problem. We apply a new multiple testing
correction appropriate for the problem which is based on analytical
results, and calculate adjusted p-values to obtain lists of enriched
GO-terms that can hint to biological interpretation and hypothesis
generation from the results of high throughput methods. In the
following we describe our framework and compare the multiple testing
correction applied here to earlier approaches. Subsequently, we
investigate the robustness of our framework with respect to the exact
gene group composition. We also show the results of our method applied
to a previously published microarray experiment: namely genes that
have been found to be cell-cycle regulated in human HeLa cell lines
\cite{Whitfield02}. Further data and a heuristical approach to
determine the family wide error rate (FWER) are discussed in the
supplement \cite{supplement}.

\section{Methods}

\begin{figure*}
\begin{center}
\begin{tabular}{cc}
\begin{minipage}[b]{7cm}
\begin{center}
\includegraphics[width=7cm]{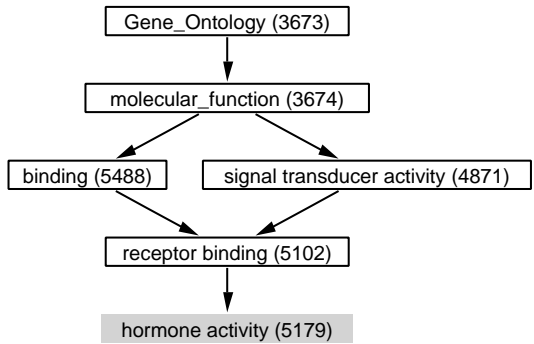}
\end{center}\end{minipage}
&
\begin{minipage}[b]{8cm}
\caption{ Part of the directed acyclic graph (DAG) representing the
Gene Ontology. Annotations are usually given as terms in different parts of the 
DAG, e.g.\ term 5179 hormone activity, implying a series of more
general terms (identifiers 5102, 5488, 4871, 3674, 3673). We consider
all levels of the GO graph and our algorithm counts for each node in
the DAG how many genes in the test set and reference set are
implicitly or explicitly annotated with this
term.\label{figure:dag} }\end{minipage}
\end{tabular}
\end{center}
\end{figure*}

\begin{figure*}
\begin{center}
\begin{tabular}{cc}
\begin{minipage}[c]{5cm}
\begin{center}
\includegraphics[width=5cm]{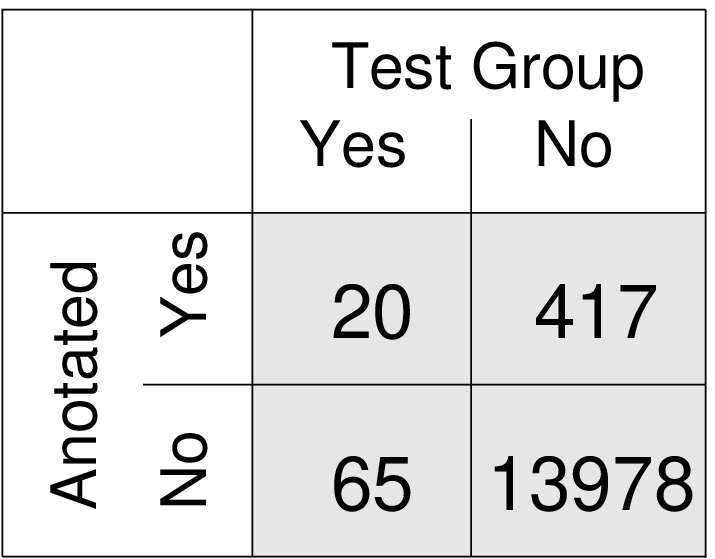}
\end{center}\end{minipage}
&
\begin{minipage}[c]{11cm}
\caption{$2\times 2$ contingency table of gene frequency that is
calculated for each term. Each gene in categorized in two ways:
whether it belongs to the test group and whether it is annotated with
the term under consideration. This figure shows the contingency table
for the term {\em mitotic cell cycle} in the group of genes that are
up-regulated during the G2-phase of the cell cycle. In total, 14480
genes are in the reference group. 437 genes are annotated with this
specific term, 20 of them are in the G2-phase gene group. 14043 are
not annotated with this term. Out of this, 65 are in the test
group. The number of genes in the test group is 85.\label{figure:contingency} }
\end{minipage}
\end{tabular}
\end{center}
\end{figure*}

\subsection{Data preparation}
In order to profile gene groups we require four data sources: a test
group of genes (e.g.\ up-regulated genes), a reference group (e.g.\ all
significantly expressed genes), GO annotations for these genes, and
the Gene Ontology. Many chip-manufacturers provide GO annotations for
the genes covered by their chips (e.g.\ Affymetrix (TM), see \cite{affy}),
otherwise gene groups can be
annotated using tools like HomGL \cite{Bluthgen04}. The current version of the Gene
Ontology can be downloaded from the website of the Gene Ontology
Consortium \cite{go}. In this analysis we used the Gene Ontology as of 11th
February, 2004. The Gene Ontology can be represented as a directed
acyclic graph (DAG) where the nodes represent the terms
\cite{Bard04}. Annotations are usually given as terms within the DAG
implying a series of more general annotations upward in the GO graph,
as illustrated in Figure \ref{figure:dag}.

\subsection{Statistical framework}

For each term in the ontology we ask whether this particular term is
enriched in the test group as compared to the reference group. To test
this we categorize each gene in two ways: first, whether it is
annotated with the term under consideration or not, and second,
whether it belongs to the test group or not. Based on these categories
we build a $2\times 2$ contingency table of gene frequencies for each
term. Figure \ref{figure:contingency} shows the structure of such a
contingency table. Using Fisher's exact test \cite{fisher}
we compute p-values that
allow to detect and quantify associations between the two
categorizations. Fisher's exact test is based on the hyper-geometric
distribution, and works in a similar way as the $\chi^2$-test for
independence. The $\chi^2$-test provides only an estimate of the true
probability values, and it is not accurate if the marginal
distribution is very unbalanced or if we expect small frequencies
(less than five) in one of the cells of the contingency table. Both
situations are typical for the task and data under
consideration. Although Fisher's exact test can in principle quantify
the reduction of a term with respect to the reference group we focus
on enrichment or association. A reduction is unlikely to be detected
in typical data sets, since the test group is usually much smaller
than the reference group.

To control the number of false discoveries, we determine adjusted
p-values to control the false discovery rate (FDR) that quantifies the
expected portion of false discoveries within the positives. If there
is no prior expectation about an association between the gene list and
any biological process, one might favor the family-wise error rate
(FWER), see \cite{supplement}. However, the typical case in
profiling gene lists is that one expects some terms to be enriched. In
this case the FDR is an adequate measure of false discoveries. Both
rates can be reliably estimated by resampling simulations, but this
method suffers from very long runtime even on modern
computers. Alternatively, several approaches exist to estimate the FDR
from the single-test p-values (e.g.\ Benjamini-Hochberg, and
Benjamini-Yekutieli \cite{Dudoit03}). These methods are designed to
cope with general problems but turn out to be not particularly suitable 
for the specific problem considered here. For the specific problem of 
profiling gene groups, the expected number of false discoveries (NFD) 
for a given p-value threshold can be determined exactly by an analytical 
expression. Consequently, we can calculate the FDR exactly.

\subsection{Number of false discoveries (NFD)}
%
The number of genes in the reference group is denoted by $N$, the test
group is a subgroup of the reference group, with $T$ being the number
of genes within this group. With $K$ we denote the number of GO-terms
which annotated genes in the reference group. We index these GO-terms
with $i=1...K$, and $Z_i$ denotes the number of genes in the reference
group being annotated by GO-term $i$. In the example of Figure
\ref{figure:contingency} these numbers would be: $N=14480$, $T=85$,
and $Z_i=457$.

For a given p-value threshold $\alpha$ we obtain the expected number
of false discoveries ($\mathrm{NFD}(\alpha)$) by summing over all
possible tests with weights according to the probability of their
occurrence
\begin{equation}
\label{eqn:1}
\langle \mathrm{NFD}(\alpha)\rangle=\sum_{i=1}^K\mathrm{Pr}(p_i\le\alpha)\, .
\end{equation}
Here $\mathrm{Pr}(p_i\le\alpha)$ denotes the probability that the
unadjusted p-value of term $i$ with its marginal distributions matches
the threshold $\alpha$ by chance. We use the hypergeometric
distribution $h(j,T,N,Z_i)$
\begin{equation}
h(j,T,N,Z_i)=\frac{Z_i!T!(N-Z_i)!(N-T)!}{N!j!(Z_i-j)!(T-j)!(N-Z_i-T+j)!}\,,
\end{equation}
to describe the probabilities of observing
$j$ annotations given the marginal distribution $(T,N,Z_i)$. Then
$\mathrm{Pr}(p_i\le\alpha)$ can be calculated by
\begin{equation}
\mathrm{Pr}(p_i\le\alpha)=\sum_j^{p_f(j,T,N,Z_i)\le\alpha}h(j,T,N,Z_i)
\;, 
\end{equation}
where $p_f(j,T,N,Z_i)$ the p-value of the one-sided Fisher test 
\cite{fisher} for $j$ or more annotations in the test group and can be calculated by
summing over the hypergeometric distribution:
\begin{equation}
p_f(j,T,N,Z_i)=\sum_{k=j}^{\mathrm{min}(Z_j,T)}h(j,T,N,Z_i) \;. 
\end{equation}
In order to validate our analytical result for $\langle \mathrm{NFD}(\alpha)\rangle$ we
estimate the number of false discoveries using resampling
simulations. We keep the reference group fixed with N genes and then
select random test groups of size $T$. The expected number of false
discoveries for a specific p-value threshold $\alpha$ is estimated by
the mean number of positive tests in the resampling runs. Note that
the correlations between terms induced by the structure of the graph
and by the annotation do not influence the mean number of false
discoveries but skew the distribution.

\subsection{False discovery rate (FDR)}

If one expects some terms to be enriched in the test group,
controlling the false discovery rate (FDR) is the appropriate method
\cite{Dudoit03}. The FDR gives an estimate of the proportion of the
expected number of false discoveries $\mathrm{NFD}(\alpha)$ among all positives
R($\alpha$) for a given p-value threshold $\alpha$: $\mathrm{FDR}(\alpha)=\langle\mathrm{NFD}(\alpha)\rangle/\mathrm{R}(\alpha)$.
Since we can calculate $\langle\mathrm{NFD}(\alpha)\rangle$ exactly,
the exact determination of $\mathrm{FDR}(\alpha)$ is possible, and an
adjusted p-value for exact control of the portion of false discoveries
is given by $p_{\mathrm{FDR}}(p)=\mathrm{min}(\mathrm{FDR}(p),1)$. Each list of
terms that fulfills the criterion $p_{\mathrm{FDR}}(p)\le 0.05$ is expected to
contain 5\% terms that are false discoveries.

\begin{figure*}
\begin{center}
\begin{tabular}{cc}
\begin{minipage}[c]{8cm}
\begin{center}
\includegraphics[width=7.5cm]{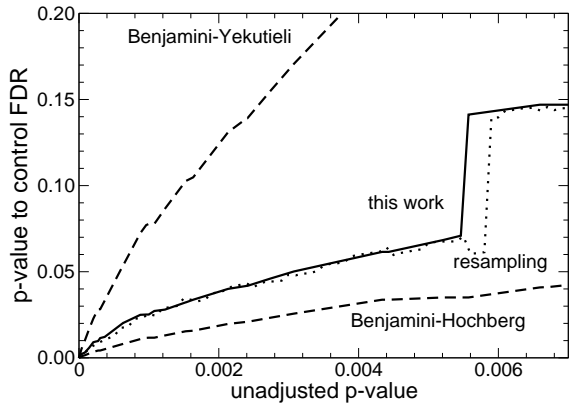}
\end{center}\end{minipage}
&
\begin{minipage}[c]{7cm}
\caption{The adjusted p-value for the group of genes expressed in the
G2-phase of the cell cycle in HeLa cells to control the FDR estimated
by our approach (solid) are in excellent agreement with the resampling
simulations (dotted). Adjusted p-values (dashed) with multiple step
Benjamini-Hochberg and Benjamini-Yekutieli are shown with dashed
lines.\label{figure:fdr} }
\end{minipage}
\end{tabular}
\end{center}
\end{figure*}

Figure \ref{figure:fdr} shows comparisons of the FDR calculated by our
methods with other approaches for data sets described below. Our
method is in excellent agreement with the adjusted p-values calculated
by resampling simulations which provide a reliable estimate of the
true FDR. In the supplement \cite{supplement} we show that analytical 
and numerical results are in excellent agreement for a wide range of 
test group sizes and various data sets. 

A particularly interesting property of this type of curve are the sudden jumps
of the FDR. These jumps are caused by the discrete values of the
p-values in the Fisher's exact test due to the discrete nature of the
contingency table. In the example discussed here, some GO-terms
annotate only two genes, and their lowest possible p-value is
0.033. Therefore these GO-terms can never be significantly enriched,
when a single-test p-value threshold is 0.03. In contrast to our
method and the resampling simulations, the methods according to
Benjamini-Yekutieli (BY) and Benjamini-Hochberg (BH) cannot account
for these discrete p-values. Therefore, they cannot reproduce the
sudden jumps and do not estimate the FDR precisely for our specific
problem. Some authors, like the authors of FatiGO \cite{AlShahrour04},
try to avoid this problem by only testing at a certain level or like
in GeneMerge \cite{Castillo-Davis03}, limit the search to terms that
annotate two or more genes. Both approaches have disadvantages: The
first approach neglects that the Gene Ontology is a directed, acyclic
graph and the concept of levels can only apply to trees. Additionally,
it is unclear whether all terms at a certain {\em level} are similarly
specific. The limitation used in the latter approach is rather
arbitrary, as it is not a priori clear, which number of genes need to
be annotated by a term that it can get significant.  The resampling
simulations in Figure \ref{figure:fdr} show that the appropriate
threshold allowing 5\% false discoveries is in this case around
p=0.003. The upper limit for the FDR provided by BY is far to
conservative: if one would allow 5\% of false discoveries in Figure
\ref{figure:fdr}, the single-test p-value threshold estimated by BY
would be around p=0.0005. Thus this method misses significant
results. The Bonferroni correction is even more conservative then BY, 
see \cite{supplement}.
The estimate provided by BH (p=0.0075) is not reliable for
the type of data under consideration, yielding more false discoveries
then specified by the threshold. Other data sets lead to similar
results concerning the performance of standard multiple testing
corrections, see \cite{supplement}.

\section{Results}

\subsection{Cell cycle regulated genes}

\begin{table*}[htbp]
\caption{Significantly enriched terms (FDR $\le 0.05$) of the five
groups G1/S, S, G2, G2/M, and M/G1 that fall into the cell cycle
categories (CCC) gap phases (G), replication (R), and mitosis (M).}
\label{table:cellcycle}
\begin{center}
\scriptsize
\begin{tabularx}{\linewidth}{p{0.5cm}r>{\raggedright}p{7cm}XXXXX}
\hline
 & & & \multicolumn{5}{c}{adjusted p-values to control FDR} \\
 & & & \multicolumn{5}{c}{for the groups according to Whitfield et.al.\ \cite{Whitfield02}} \\
CCC & ID & GO term & G1/S & S & G2 & G2/M & M/G1\\
\hline
G & 8156 & negative regulation of DNA replication & $8\cdot 10^{-5}$ &  &  &  &\\
R & 5659 & delta DNA polymerase complex & $5\cdot 10^{-4}$ &  &  &  &\\
R & 6270 & DNA replication initiation & $2\cdot 10^{-4}$ &  &  &  &\\
G & 45003 & double-strand break repair via synthesis-dependent strand annealing & $1\cdot 10^{-3}$ &  &  &  &\\
G & 45002 & double-strand break repair via single-strand annealing & $1\cdot 10^{-3}$ &  &  &  &\\
G & 731 & DNA repair synthesis & $1\cdot 10^{-3}$ &  &  &  &\\
R & 42575 & DNA polymerase complex & $1\cdot 10^{-3}$ &  &  &  &\\
G & 724 & double-strand break repair via homologous recombination & $6\cdot 10^{-3}$ &  &  &  &\\
R & 3891 & delta DNA polymerase activity & $6\cdot 10^{-3}$ &  &  &  &\\
G & 725 & recombinational repair & $6\cdot 10^{-3}$ &  &  &  &\\

R &  3887 & DNA-directed DNA polymerase activity &  $1\cdot 10^{-2}$ & & & & \\
G &  726 & non-recombinational repair &  $2\cdot 10^{-2}$ & & & & \\
R &  6312 & mitotic recombination &  $5\cdot 10^{-2}$ & & & & \\
R  & 84 & S phase of mitotic cell cycle & $1\cdot 10^{-8}$ & $3\cdot 10^{-7}$  &  &  &\\
R & 6260 & DNA replication & $1\cdot 10^{-8}$ & $3\cdot 10^{-7}$ &  &  &\\
R & 30894 & replisome & $6\cdot 10^{-3}$ & $1\cdot 10^{-5}$ &  &  &\\
R & 5657 & replication fork & $9\cdot 10^{-3}$ & $2\cdot 10^{-5}$ &  &  &\\
R & 6261 & DNA dependent DNA replication & $1\cdot 10^{-8}$ & $7\cdot 10^{-4}$ &  &  &\\
G & 6974 & response to DNA damage stimulus &  $1\cdot 10^{-2}$ & $5\cdot 10^{-3}$ &  &  &\\
G & 6281 & DNA repair & $4\cdot 10^{-3}$ & $9\cdot 10^{-3}$ &  &  &\\
R & 3896 & DNA primase activity &  & $5\cdot 10^{-3}$ &  &  &\\
R & 4748 & ribonucleoside-diphosphate reductase activity &  & $7\cdot 10^{-3}$ &  &  &\\
R & 5658 & alpha DNA polymerase:primase complex  &  & $7\cdot 10^{-3}$ &  &  &\\
R &  6269 & DNA replication, priming & &  $1\cdot 10^{-2}$ & & & \\
R &  5663 & DNA replication factor C complex  & &  $3\cdot 10^{-2}$ & & &\\
M & 7017 & microtubule-based process &  &  & $2\cdot 10^{-5}$ &  &\\
M & 70 & mitotic chromosome segregation &  &  & $6\cdot 10^{-5}$ &  &\\
M & 5874 & microtubule &  &  & $1\cdot 10^{-4}$ &  &\\
M & 90 & mitotic anaphase &  &  & $2\cdot 10^{-4}$ &  &\\
M & 30705 & cytoskeleton-dependent intracellular transport &  &  & $4\cdot 10^{-4}$ &  &\\
M & 7018 & microtubule-based movement &  &  & $4\cdot 10^{-4}$ &  &\\
M & 7093 & mitotic checkpoint &  &  & $2\cdot 10^{-3}$ &  &\\
M & 7059 & chromosome segregation &  &  & $2\cdot 10^{-3}$ &  &\\
M & 7010 & cytoskeleton organization and biogenesis &  &  & $9\cdot 10^{-3}$ &  &\\
M & 30261 & chromosome condensation &  &  & $9\cdot 10^{-3}$ &  &\\
M & 45298 & tubulin &  &  & $1\cdot 10^{-2}$ &  &\\
M & 88 & mitotic prophase &  &  & $2 \cdot 10^{-2}$&  &\\
M & 7076 & mitotic chromosome condensation &  &  & $2\cdot 10^{-2}$ &  &\\
M & 5813 & centrosome &  &  & $3\cdot 10^{-2}$ &  &\\
M & 922 & spindle pole &  &  & $4\cdot 10^{-2}$ &  &\\
M & 5815 & microtubule organizing center &  &  & $4\cdot 10^{-2}$ &  &\\
M & 7067 & mitosis &  &  & $9\cdot 10^{-9}$ & $6\cdot 10^{-7}$ &\\
M & 87 & M phase of mitotic cell cycle &  &  & $9\cdot 10^{-9}$ & $7\cdot 10^{-7}$ &\\
M & 15630 & microtubule cytoskeleton &  &  & $1\cdot 10^{-5}$ & $2\cdot 10^{-6}$ &\\
M & 280 & nuclear division &  &  & $9\cdot 10^{-9}$ & $2\cdot 10^{-6}$ &\\
M & 279 & M phase &  &  & $9\cdot 10^{-9}$ & $3\cdot 10^{-6}$ &\\
M & 5819 & spindle &  &  & $1\cdot 10^{-2}$ & $1\cdot 10^{-5}$ &\\
M & 775 & chromosome, pericentric region &  &  & $2\cdot 10^{-2}$ & $9\cdot 10^{-3}$ &\\
M & 5856 & cytoskeleton &  &  & $9\cdot 10^{-3}$ & $3\cdot 10^{-2}$ &\\
M & 7088 & regulation of mitosis &  &  & $1\cdot 10^{-4}$ & $1\cdot 10^{-2}$ &\\
M & 910 & cytokinesis &  &  & $5\cdot 10^{-6}$ & $2\cdot 10^{-2}$ &\\
\hline
\end{tabularx}
\end{center}
\end{table*}

Whitfield et al.\ \cite{Whitfield02} analyzed time series
of gene expression profiles of cell-cycle synchronized HeLa cells
leading to groups of genes exhibiting peak expression in specific
phases of the cell cycle. Their analysis yielded 874 different genes
which were assigned to five groups of cell cycle regulated genes by
their correlation to an idealized expression profile generated from
several well studied marker genes for G1/S-phase, S-phase, G2-phase,
G2/M-phase and M/G1-phase. The functions of the genes proposed by this
assignment procedure were found to match published literature and
other reports of cell cycle gene expression experiments. In addition,
assignment of GO terms was performed.  A good correlation between the
cell cycle phase when peak expression of a gene was observed and its
assigned GO terms was observed. However, no statistical analysis of
this correlation was performed by the authors.

In the following we applied our method to these five published gene
groups first to determine automatically the GO terms which are
significantly overrepresented in each of the five different gene
groups, and second to find out whether these overrepresented GO terms
match the groups function assigned by Whitfield et
al. \cite{Whitfield02}. Using HomGL \cite{Bluthgen04} we could
annotate 15910 of 37137 genes on their chip with at least one
GO-Terms. This set was used as the reference group.  To reflect our
prior expectation that some cell-cycle related terms are enriched, we
used the adjusted p-value to control the FDR with a threshold of 5\% as
the criterion for significance. Between 1 and 51 GO terms (mean: 28)
were found to be significant (the detailed lists of all terms and
figures displaying the significant terms in the DAG can be found in
the supplement \cite{supplement}. That only one
significant term in the M/G1 group was observed is not surprising as
Whitfield et al. \cite{Whitfield02} assigned here genes to the group which are
expressed during the physical act of mitosis and which are persisting
into G1 phase. These genes belong to processes such as cell adhesion
or membrane trafficking which are not necessarily cell cycle specific
and fall into a wide range of GO categories which therefore are not
significantly overrepresented.

To reach our second aim, we categorize all GO terms which are
significantly enriched in at least one of the groups into four
categories with respect to the cell-cycle phases: gap phases (G),
replication (R), mitosis (M), and unspecific (U). Subsequently, we
order the GO terms which fall into the cell cycle relevant categories
gap phases, replication and mitosis with respect to the groups of
Whitfield et al.\ \cite{Whitfield02}. The result is shown in Table
\ref{table:cellcycle}. Through this procedure, we can examine to which degree cell cycle
categories match the groups by Whitfield et al. The most prominent
finding is that all GO terms of the replication category stem from the
G1/S or S groups and none of these terms originates from either the
G2/M, M, or M/G1 groups. Similarly, all terms of the mitosis category
originate from the G2, or G2/M groups and none from G1/S or S
group. Seven of the terms which fall into the gap phase category
originate from the G1/S group and another two terms are contained in
the significant terms of the G1/S and S-phase group, while terms from
all groups are found in the unspecific cell cycle group (see
Table 1 of \cite{supplement}). In conclusion we can confirm that the terms
that are significant in our analysis provide a reasonable biological
characterization of the gene groups suggest by Whitfield et
al.\ \cite{Whitfield02}. We can clearly separate mitosis and replication. In addition,
our analysis indicates the close relationship of G2- and M-phase, and
G1- and S-phase respectively. This example shows that our method can
provide useful functional characterization of a given gene group
without using prior knowledge. For further examples refer to the
supplement \cite{supplement}.

\subsection{Robustness of biological profiles}

\begin{figure*}
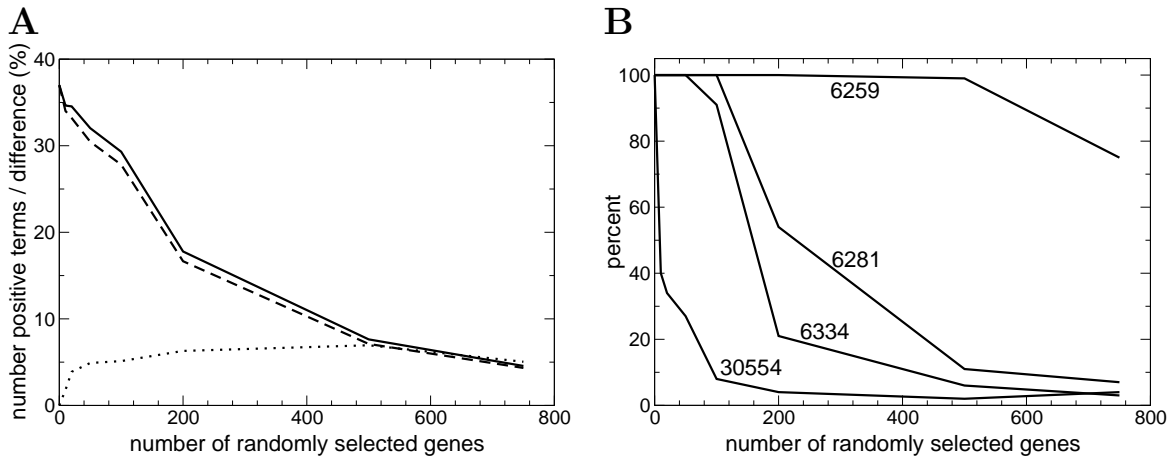

\begin{center}
\begin{tabular}{ll}
 \Large \bf A & \Large \bf B \\
\includegraphics[width=7.5cm]{figure4a.eps}
&
\includegraphics[width=7.5cm]{figure4b.eps}
\end{tabular}
\end{center}
\caption{ The effect of adding randomly selected genes to the test
group: to the gene group of cell-cycle regulated genes expressed in
the S-phase (221 genes), we add randomly 10, 20, 50, 100, 200, 500 and
750 genes from the reference group and profile these gene groups. (A)
The solid line shows the mean number of significant terms in 100
runs. The dashed line displays the number of those terms, which are
significant in the initial group. The dotted line shows the difference
between both lines in percent. (B) Percentage of runs where the terms
with identifier 30554, 6334, 6281 and 6259 are still detected as
significant.
\label{figure:robustness}}
\end{figure*}

When extracting a gene group from high-throughput experiments, one
always has to deal with the trade-off between specificity and
sensitivity. By increasing the group size, an increased portion of
genes is included in the group just by chance and not due to
biological or functional reasons. Therefore it is often not clear
which genes to include in or exclude from a certain gene group,
e.g.\ by choosing a specific threshold. In principle these falsely
assigned genes might result in falsely discovered GO terms.

In the following we show that the results of our method do not depend
critically on the precise composition of the test group, and are
robust with respect to a high portion of genes which are accidentally
assigned to the group. Here we assume that the falsely assigned genes
are randomly distributed on the microarray. This assumption does not
necessarily hold for gene lists obtained with rather naive microarray
analysis, since e.g.\ the absolute signal or cross hybribridization
might induce a bias. However, there are approaches to limit such a
bias, for example by estimating the variance for each gene from
replications \cite{Dudoit02}, or by variance stabilization
\cite{Huber02}. 
We address the robustness by profiling the gene group of 221
cell-cycle regulated genes expressed in the S-phase
\cite{Whitfield02}, and adding randomly selected genes to this
group. Initially, 37 terms are reported to be significant
(FDR$<$0.05). Subsequently 10, 20, 50, 100, 200, 500, and 750 randomly
selected genes from the reference group are added to
the initial test group and the resulting groups are profiled. We
repeated this procedure 100 times and determined how much of the
initial profile is preserved. Figure \ref{figure:robustness}A shows
that the number of significantly enriched terms decreases if we add
more and more random genes. However, the profile is remarkable robust:
even if we add 200 randomly selected genes, we can still detect 17 of
the initially 37 significant terms. Interestingly, from the difference
between the solid and dashed line we see that only about 5\% of the
terms in the resulting profiles are not contained in the initial
profile for the S-phase, regardless of the number of randomly selected
genes added. These are potentially falsely discovered terms confirming
our criterion of the adjusted p-value to control the
FDR$<$0.05. Figure \ref{figure:robustness}B shows the robustness of
terms with different initial FDR. As expected, terms with a FDR just
below the threshold of 0.05, like adenyl nucleotide binding (id 30554,
FDR=0.044), are unlikely to be detected after adding many randomly
selected genes. However, terms with intermediate significance, like
nucleosome assembly (id 6334, FDR=0.0091), and DNA repair (id 6281,
FDR=0.00099), can still be detected nearly always after adding 100
randomly selected genes. Remarkably, highly significant terms like DNA
metabolism, (id 6259, FDR=$5.2 \cdot 10^{-8}$), are found in 99\% of all
cases even after adding 500 randomly selected genes. These results
show that our framework performs reasonably well even if many randomly
selected genes are in the gene groups. Especially intermediate and
highly significant terms will persist. Furthermore, the framework
controls the number of falsely discovered terms reliably.  

\section{Discussion}

In this study we propose a statistical framework to find molecular
functions, biological processes and cellular locations significantly
associated with gene groups. Our approach allows an unbiased
biological profiling of gene groups beyond the single gene level. We
pay special attention to the multiple testing problem, and we give an
exact determination of the expected number of false discoveries. We
validate our framework with resampling simulations, and find that we
can calculate adjusted p-values as reliable as with resampling
simulations. Adjusted p-values can differ from the single test
p-values by a factor of more than $10^4$. In contrast to resampling
simulations, which are very slow (up to hours), our approach needs
just a few seconds. Comparison with our approach shows that the method
of Benjamini-Yekutieli is too conservative, resulting in adjusted
p-values which are typically 2-6 times higher. Thus, this methods would miss
many significant results. The Benjamini-Hochberg estimate is not
reliable for the problem under consideration.

Furthermore, we show that our method is robust with respect to
randomly assigned genes in the gene groups. Therefore the biological
profiles do not depend critically on the details of the prior analysis
yielding the groups, e.g.\ threshold, parameters of cluster analysis
methods, and normalizations. For the cell-cycle data of Ref.\
\cite{Whitfield02} we demonstrate that the profiles allow a reasonable
biological interpretation of the gene groups. No clearly irrelevant
terms are reported. The fact that we can separate mitosis and
replication confirms the initial grouping by Whitfield et al.\
\cite{Whitfield02} Therefore, we conclude that our framework is indeed
sensitive enough to find the relevant biological profiles in gene
groups.

Additionally, our framework has been applied to predict the functional
targets of transcription factors by profiling gene lists that display
clusters of binding sites in their upstream regions
\cite{Kielbasa04,Bluthgen05}. Although the gene groups that display
clusters of binding sites are dominated by non-functional sites, the
rigorous statistics allows to make reliable predictions for the true
functional targets. The example studied in this paper and in the
supplement \cite{supplement} illustrate that the
combination of high-throughput technologies, Gene Ontology, and a
careful statistical analysis can automate the complex and laborious
task of understanding the function of gene groups.

There are several software implementations available to profile gene
groups using Gene Ontology, including Onto-Express \cite{Draghici03},
EASE+David \cite{Hosack03,Dennis03}, GoSurfer \cite{Zhong03}, GoMiner
\cite{Zeeberg03}, GeneMerge \cite{Castillo-Davis03}, FatiGO
\cite{AlShahrour04}, and GOstat \cite{Beissbarth04}. The software
packages listed before, with the exception of Onto-Express (free
version) and GoMiner, have some multiple testing correction. We
conclude from our studies, that results of applications that do not
use multiple testing corrections (Onto-Express and GoMiner) are hard
to be interpreted since here false positive predictions will
dominate. On the other hand, the packages using appropriate standard
multiple testing corrections (Bonferroni in Gene Merge,
Benjamini-Yekutieli in FatiGo, GoStat, GOSurfer) do give control of
the number of false discoveries, but are too conservative, and
therefore have less power. Only EASE+David uses a jackknife procedure
similar to resampling to correct for multiple testing, which can give
more robust scores, although they cannot be interpreted as adjusted
p-values. A software package (GOSSIP) for profiling gene groups that 
uses our method and visualizes the results is made freely available at
\cite{supplement}.

\section{Acknowledgements}

 We thank Szymon Kielbasa, Johannes Schuchhardt, Frank Kleinjung,
Martina Geheeb, Steffen Grossmann and Matthias Futschik for fruitful
discussions and comments on the manuscript. NB and BC acknowledge
support by the Deutsche Forschungsgemeinschaft (SFB 618), MS is
supported by the BMBF. The work of DB has been supported by the BMBF 
grant number 01GR0444.
\bibliographystyle{plain}
  \bibliography{gossip}


\end{document}